\documentstyle[prc,aps,epsfig,multicol]{revtex}
\draft
\normalbaselines
\def\bm {\bibitem}
\def\be {\begin{equation}}
\def\ee {\end{equation}}
\def\nn {\nonumber}
\def\bea {\begin{eqnarray}}
\def\eea {\end{eqnarray}}

\def\mr {M_R}
\def\kb {|\vec k|}

\def\mn {m^{*}_n}
\def\kb {\bf k}
\def\2l {\frac{{f_i}}{(2\lambda + 1)}}
\def\qs {q\!\!\!/}
\def\ks {k\!\!\!/}
\def\ps {p\!\!\!/}

\begin{document}

\title{Aspects of meson properties in dense nuclear matter}
\author{ 
Octavian Teodorescu, 
Abhee K. Dutt-Mazumder\footnote{Current address: TRIUMF Theory Group, 4004 Wesbrook Mall, Vancouver, BC, Canada V6T 2A3} Charles Gale}
\address{McGill University, Montreal, Quebec, Canada }
\maketitle

\begin{abstract}
We investigate the modification of meson spectral 
densities in dense nuclear matter at zero temperature. These effects are 
studied in a fully relativistic mean field model which goes beyond the 
linear density approximation and also includes baryon 
resonances. In particular, the role of $N^*(1520)$ and $N^*(1720)$ on the 
$\rho$ meson spectral density is highlighted. 
Even though the nucleon-nucleon loop and 
the nucleon-resonance loop contribute with the opposite sign, an overall 
reduction of $\rho$ meson mass is still observed at high density.  
Importantly, it is shown that the
resonances cause substantial broadening of the $\rho$ meson spectral density
in matter and also induces non-trivial momentum dependence.
The spectral density of the $a_0$ meson is also shown.
We study the dispersion relations and collective
oscillations induced by the $\rho$ meson propagation in nuclear matter
together with the influence of the mixing of $\rho$  with
the $a_0$ meson. 
The relevant expression for the plasma frequency is also recovered 
analytically in the appropriate limit.
\end{abstract}
\vspace{0.3 cm}


\section{Introduction}

Electromagnetic radiation constitutes a privileged probe of the properties of matter under extreme conditions. This owns partly to the fact that it decouples from the strongly interacting system without significant rescattering and also because the virtual photons enjoy a direct coupling to vector mesons.
Lepton pairs carry thus valuable information about the in-medium properties. 

Among the light vector mesons, the $\rho$ acquires a special importance because of its large 
decay width. Therefore, this might serve as a chronometer 
and thermometer to probe temporarily produced hot and dense 
hadronic matter. Even though $\omega$ or $\phi$ mesons do not have 
this desired short lifetime, in the medium they might  
undergo sufficient broadening leading to a shorter lifetime \cite{weise}. For the present 
purpose, however, we first mostly concentrate on the $\rho$ meson and we discuss the scalar-isovector sector later on.

The in-medium properties of the $\rho$ meson have been estimated in variety
of models ranging from QCD sum-rules \cite{abhee01} to chiral models like 
Nambu-Jona Lasinio, effective hadronic Lagrangian approaches
\cite{shiomi94} and mean-field models. It is fair to say that, at this point, a clear consensus is still lacking but that important progress has been realized over the past few years. For a review see \cite{Rappreview}. The angle of this work consists of uniting several physical aspects we find important but that had not been treated together in a unique approach. So here, the $\rho$ spectral density
in dense nuclear matter is studied and the importance of the $N^*$(1520) and 
$N^*$(1720) is reiterated, but a relativistic calculation is performed and we go beyond the linear density approximation (LDA). A relativistic calculation has recently been presented in \cite{post01} in the LDA.  We show that 
LDA is a good approximation for densities below nuclear matter
density, but for higher densities multiple scattering becomes important.
We report a quantitative comparison of the results obtained in the linear 
approximation with a resummed one loop calculation. We also incorporate
the effect of interacting nuclear matter  through the
scalar and vector meson mean fields, motivated by the Walecka model. 
The role of resonances and that of the nucleon loops are examined separately. 
Finally, we include the recently discussed mixing effects \cite{teodorescu00,teodorescu01}, 
and report on the spectral density of the $a_0$ for the first time.

The paper is organized as follows. First the formalism is outlined followed
by a discussion of the $\rho$ meson properties involving nucleons. Then
we consider the effect of the resonances on the in-medium spectral
densities. Later we discuss the effect of mixing. We also present the 
spectral density of the $a_0$ meson which supplements our understanding of 
the mixed 
propagator of the $\rho$ in nuclear matter. The calculations are done
in a fully relativistic formalism including the effect of the mean field. At
places the mathematical details are relegated to the appendix. Finally
we discuss the results and conclude.

\section{Formalism}

It is well known that the spectral density is actually the imaginary part 
of the propagator which in turn is
related with the polarization functions of the meson. Therefore, 
we first discuss the properties of the $\rho$ meson polarization function 
in dense nuclear matter. 

Essentially, the spectral density is related with the collective excitation
induced by the $\rho$ meson by its propagation in nuclear matter. This is
analogous to the photon propagation in QED plasma where the propagating
particle picks up the collective modes from the system arising out of the
density fluctuation. This is commonly known as plasma oscillation. Even
though in the present work the main focus is not to recover the characteristic
features of the plasma oscillation induced by the $\rho$ meson, nevertheless,
we outline the formalism {\em \`a la} Chin\cite{chin77}
in order to be able to discuss
the spectral density in terms of the dielectric response function of the
nuclear matter. This enables us to incorporate the effect of meson mixing
in a straightforward manner \cite{saito98}.

The $\rho$ meson, being a massive spin one particle, can have both longitudinal
and transverse excitations depending upon whether its momenta is perpendicular
or parallel to the spin. Furthermore, in matter the Lorentz symmetry is
broken and these two modes will have different characteristic features. These
states are designated as $\Pi^L(q_0,|{\bf q}|)$, $\Pi^T(q_0,|{\bf q}|)$ 
with L and T denoting longitudinal and transverse modes.

As already mentioned, we consider the coupling of $\rho$ meson with n-n,
n-R and $\pi$-$\pi$ states and therefore what we have is the following
\bea
\Pi^{L(T)}= \Pi_{nn}^{L(T)}+ \Pi_{Rn}^{L(T)}+ \Pi_{\pi\pi}^{L(T)},
\eea
with $R=N^*(1520), N^*(1720)$. First we present a general formalism without
the effect of mixing and later we shall address the issue of the possible
mixing and of the corresponding modifications.

To describe the nuclear matter ground state we invoke the mean
field approach of quantum hadrodynamics and consequently the effective
nucleon mass is generated through the $\sigma$ meson tadpole of scalar
mean field potential \cite{walecka86}. 
The nucleon mass is determined by solving the
following equation self-consistently.
\bea
m_n^* = m_n -4(\frac{g_\sigma}{m_\sigma})^2\int_0^{k_F}
\frac{d^3k}{(2\pi)^3}\frac{m_n^*}{\sqrt{k^2+m_n^{*2}}} .
\eea

To study the collective excitation of the system, the relevant quantity
is the dielectric function which actually characterizes the eigenvalue
condition for the collective modes. In the language of field theory 
this is equivalent to solving the Schwinger-Dyson equation to determine
the dressed propagator. The relevance
of summing over the ring diagrams for the study of vector meson propagation
is discussed at length in Ref. \cite{chin77}. 

The vector meson propagation
is calculated by summing over ring diagrams, a diagrammatic equivalent
of the random phase approximation (RPA), which consists of repeated
insertions of the lowest order polarization,
as illustrated in Fig.\ref{fig:rings}  \cite{chin77} below.
\begin{figure} [htb!]
\begin{center}
\epsfig{file=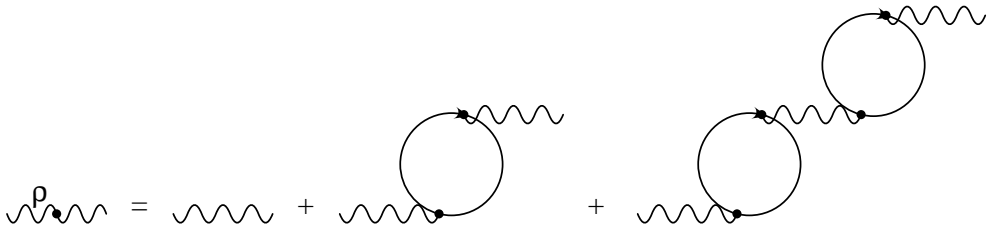,height=3.0cm,angle=0}
\end{center}
 \caption[Random Phase Approximation]
  {\small Ring diagrams relevant for the random phase approximation. 
\label{fig:rings}}
\end{figure}

We make use of the Dyson equation to carry out the summation
\begin{equation}
D_{\mu\nu}(q)= D^0_{\mu\nu}(q) + D^0_{\mu\alpha}(q)\Pi^{\alpha\beta}
(q)D_{\beta\nu}(q) .
\end{equation}

\noindent The poles are found from the equation
\begin{equation}
det[\delta^\nu_\mu - D^0_{\mu\alpha}\Pi^{\alpha\nu}]=0 .
\end{equation}
The bracketed term is nothing but the dielectric tensor of the system
\begin{equation}
\epsilon^\nu_\mu =\delta^\nu_\mu -D^0_{\mu\alpha}\Pi^{\alpha\nu}
\end{equation}
the determinant of which, denoted later by $\epsilon (q)$,
 is the dielectric function.
The eigenconditions for collective modes can now be expressed
as $\epsilon(q)=0$. The relevance of the set of ring diagrams and the
origin of such an eigencondition can be understood from linear
response theory where the fluctuation of the current density, the
source term for the meson field in nuclear matter, 
is ``picked up'' by the vector field.

 For later convenience we define longitudinal and transverse
dielectric functions as
\begin{eqnarray}
\epsilon_L(q)=(1+D^0\Pi_{00})(1-D^0\Pi_{33}) + D^0\Pi_{03}D^0\Pi_{30}\\
\epsilon_T(q)=1-D^0\Pi_{11}=1-D^0\Pi_{22}=1-D^0\Pi_T .
\end{eqnarray}
Here, the modified $\Pi$ functions are used and the Dirac part has also
been incorporated. $D_0=1/(q^2-m_v^2)$ is the free vector meson propagator 
of mass $m_v$. The eigenmodes of the collective oscillations are given by
\begin{equation}
\epsilon(q)=\epsilon_T^2(q)\epsilon_L(q)=0 ,
\end{equation}
corresponding to the degrees of freedom of a massive vector
particle. The two identical (or degenerate) transverse collective modes are
each given by
\begin{equation}
\epsilon_T(q)=0 \label{epsonT}
\end{equation}
and the single longitudinal mode by
\begin{equation}
\epsilon_L(q)=0 \label{epsonL} ,
\end{equation}
which yield the relevant dispersion curves. 

\subsection{Pion-pion loop}
It is well-known that the free space decay width of the $\rho$ meson
is dominantly determined by the two-pion channel. In other words its
coupling to $\pi-\pi$ loops determines the shape of the free space
$\rho$ spectral density.

The interaction between a neutral vector meson and the pions are 
given by 
\bea
{\cal L}= g_{\rho\pi\pi}(\pi\times\partial_\mu\pi)\cdot\rho^\mu .
\eea

\begin{figure} [htb!]
\begin{center}
\epsfig{file=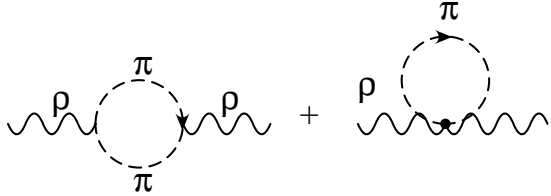,height=3.0cm,angle=0}
\end{center}
\caption{$\rho-\pi\pi $ Loop.}
\end{figure}

The real and imaginary part of the pion-pion loop have been discussed at
length in many places \cite{gale91,eletsky99}, we just quote the results here:

\bea
Re\Pi^{L(T)}=\frac{g_{\rho\pi\pi}^2 M^2}{48 \pi^2}
[(1-\frac{4m_\pi^2}{M^2})^{3/2}
ln|\frac{1+ \sqrt{1-\frac{1-4m_\pi^2}{M^2}}}{
1-\sqrt{1-\frac{1-4m_\pi^2}{M^2}}}|
+8m_\pi^2(\frac{1}{M^2}-\frac{1}{m^2_\rho}) -2 (\frac{q_0}{\omega_0})^3 
ln(\frac{\omega_0+p_0}{m_\pi})] ,
\label{realpif}
\eea

\bea
Im\Pi_\rho^{L(T)}=-\frac{g_{\rho\pi\pi}^2 M^2}{48\pi} 
(1-\frac{4 m_\pi^2}{M^2})^{3/2} . \label{impif}
\eea

The free-space spectral density of $\rho$ meson is given by the
following expression:
\bea
S_\rho(q^2)=\frac{1}{\pi}
\frac{Im\Sigma_\rho(q^2)}{(q^2-m_\rho^2-\Sigma_\rho)^2 
+ Im\Sigma_\rho^2} .
\label{freespec}
\eea
In vacuum, the above is a Lorentz invariant quantity
and a function of $q^2$. In matter, however, this breaks down and we shall 
have non-degenerate spectral densities for the longitudinal and transverse
mode of the $\rho$ meson.

\subsection{Nucleon-nucleon loop}

The $\rho$-nucleon interaction Lagrangian may be written as
\bea
{\cal L}_{int} = g_{\rho} [{\bar{N}} \gamma _\mu \tau
 N + i\frac{\kappa _\rho}{2M}{\bar{N}}
         \sigma_{\mu\nu}\tau^\alpha N\partial^\nu]
\rho^\mu_\alpha
\eea

\begin{figure} [htb!]
\begin{center}
\epsfig{file=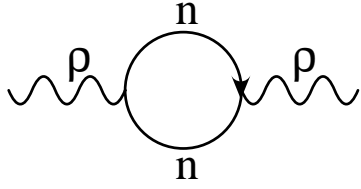,height=3.0cm,angle=0}
\end{center}
\caption{$\rho-nn$ Loop.}
\end{figure}

\bea
\Pi_{\mu\nu}^{\alpha\beta}={-i} g_V^2 S_I
\int \frac{{d^4}k}{(2\pi)^4}{\rm Tr}[i\Gamma^\alpha_\mu iG(k+q)
{i\bar\Gamma^\beta_\nu} iG(k)]
\label{nnloop}
\eea
where $S_I$ is an isospin factor ($S_I=2$ for symmetric nuclear matter). 
The vertex for $\rho$-nn is
\bea
\Gamma_\mu=\gamma_\mu - \frac{\kappa_\rho}{2m_n}\sigma_{\mu\nu}q^\nu\ .
\label{vertex}
\eea

In Eq. (\ref{nnloop}), $G(k)$ is the in-medium nucleon propagator
given by \cite{sewal}
\bea G(k_0,|{\vec k}|) = G_F(k) + G_D(k_0,{\vec k})
\label{nuclprop}
\eea

with
\bea
G_F(k)=\frac{(\ks+\mn)}{k^2-m_n^{* 2} + i\epsilon}
\label{gf}
\eea
and
\bea
G_D(k_0,|k|)=(\ks+\mn)\frac{i\pi}{E_k} \delta(k_0-E_k) \theta(k_F - |\kb|) .
\label{gd}
\eea
 
The first term in $G^0(k)$, namely $G^0_F(k)$,
is the same as the free propagator of a spin $\frac{1}{2}$ particle
while the second part, $G^0_D(k)$,
involving $\theta(k_F-|\vec k|)$,
arises from Pauli blocking and describes the modifications
brought about in nuclear matter at zero temperature. It
deletes the on mass-shell propagation of the nucleon in nuclear
matter with momenta below the Fermi momentum. It should be noted that
here, unlike the situation obtained in usual field theory in the absence of
extraneous matter,
the standard destruction operator operating on the ground state
does not give zero as long as
the momentum lies below the Fermi momenta simply because there are particles
with such momenta in the ground state. On the
other hand, particles can only be created
with a momentum higher than $k_F$ (because of Pauli blocking).
This clearly is the origin of the
second term in the propagator which has a non-covariant form. Actually,
the in-medium nucleon propagator is defined in a preferred frame, 
the rest frame for nuclear matter,
and this constitutes another important difference with usual field theory.
The Fermi momenta which plays such a central role is of course related
to the density of baryons through $\rho_{B}=(2k_{F}^{3})/(3\pi^2)$.

When calculating the polarization function (\ref{nnloop}) with the nucleon propagator (\ref{nuclprop}), there will be terms containing
``$G_FG_F$", ``$G_FG_D + G_DG_F$" and ``$G_DG_D$". 
The first term accounts for the free part {\em i.e.} the
contributions of the Dirac vacuum ($\Pi^F_{\mu\nu}$), while the rest provides
the density dependent part of the polarization $(\Pi^D_{\mu\nu})$, 
and we can write
\bea
\Pi_{\mu\nu}(q)& =& \Pi_{\mu\nu}^F(q) + \Pi_{\mu\nu}^D(q)  \\
\Pi_{\mu\nu}^F(q)& =&\frac{-i}{(2\pi)^4} g_V^2 S_I 
\int {d^4}k {\rm Tr}[\Gamma_\mu G_F(k+q)
{\bar\Gamma_\nu} G_F(k)] \\
\Pi_{\mu\nu}^D(q)& =&\frac{-i}{(2\pi)^4} g_V^2 S_I 
\int {d^4}k {\rm Tr}[\Gamma_\mu G_F(k+q)
{\bar\Gamma_\nu} G_D(k) \\ 
&+&\Gamma_\mu G_D(k+q)
{\bar\Gamma_\nu} G_F(k) + \Gamma_\mu G_D(k+q)
{\bar\Gamma_\nu} G_D(k) ] .\\
\eea

The free part of the $\rho$ self-energy 
denoted by $\Pi_{\mu\nu}^F$ is divergent and therefore needs to be
regularized. We used the dimensional regularization scheme with the
following condition:
\bea
\partial^n\Pi^F(q^2)/\partial(q^2) ^n\vert _{M^\ast_n\rightarrow M,
q^2=m_s^2}=0 (n=0,1,2...,\infty ) . 
\eea

The real part of the density dependent piece of the polarization is given by
\begin{eqnarray}
\Pi^D_{\mu\nu}
= \frac{g_v^2\pi S_I}{(2\pi)^4}\int\frac{{d^4}k}{E^{\ast}(k)}
\delta (k^0-E^{\ast}(k))\theta
(k_F-\mid \vec k\mid)
\nonumber\\
\times \Big[\frac{{\it {T}}_{\mu\nu}(k-q,k)}{(k-q)^2-M^{\ast 2}}
+\frac{{\it {T}}_{\mu\nu}(k,k+q)}{(k+q)^2-M^{\ast 2}}\Big] .
\end{eqnarray}

Beside this there is another term which involves two $\theta(k_F-|k|)$
function. That becomes operative beyond 2 times the Fermi energy\cite{chin77}.

The trace involved in the calculation of the loop diagram has three parts corresponding to vector-vector, vector-tensor and tensor-tensor terms in the vertex function (\ref{vertex}). These can be cast into the following form
\begin{equation}
T_{\mu\nu} = T^{vv}_{\mu\nu}(k,k+q)
+ T^{vt+tv}_{\mu\nu}(k,k+q)+ T^{tt}_{\mu\nu}(k,k+q)
\end{equation}
\begin{equation}
 T_{\mu\nu}^{vv}(k,k+q)= 4[k_\mu (k+q)_\nu +
(k+q)_\mu k_\nu -k\cdot (k+q)g_{\mu\nu} + M^{\ast 2}g_{\mu\nu}]
\end{equation}
\begin{equation}
 T^{vt+tv}_{\mu\nu}(k,k+q)= 4M^\ast\frac{\kappa_v}{M}q^2Q_{\mu\nu}
\end{equation}
\begin{equation}
T^{tt}_{\mu\nu}(k,k+q)=16(\frac{\kappa_v}{4M})^2
[Q_{\mu\nu}\Big ( 2(k\cdot q)^2 -q^2k^2 +
q^2(k\cdot q)-q^2M^{\ast 2}\Big )-2q^2K_{\mu\nu}] .
\end{equation}

Hence the self energy can be written as
\begin{equation}
\Pi_{\mu\nu}^D(q) = \Pi_{\mu\nu}^{vv}(q) + \Pi_{\mu\nu}^{vt+tv}(q) +
\Pi_{\mu\nu}^{tt}(q)
\label{pitt}
\end{equation}
The $\Pi_{\mu\nu}^D(q)$ functions in this case are as follows
\bea
\Pi_{\mu\nu}^{vv}&=&
\frac{g_v^2}{\pi ^3}S_I\int_0^{k_F}\frac{{d^3k}}{E^{\ast}(k)}
\frac{{\cal K}_{\mu\nu}q^2-Q_{\mu\nu}(k\cdot q)^2}{q^4-4(k\cdot q)^2}
\label{vv} \\
\Pi_{\mu\nu}^{vt+tv}&=&\frac{g_v^2}{\pi ^3}S_I
(\frac{\kappa M^\ast}{4M})2q^4Q_{\mu\nu}\int_0^{k_F}\frac{{d^3k}}{E^{\ast}(k)}
\frac{{1}}{q^4-4(k\cdot q)^2}
\label{vttv} \\
\Pi_{\mu\nu}^{tt}&=&-\frac{g_v^2}{\pi ^3}S_I
(\frac{\kappa}{4M})^2(4q^4)\int_0^{k_F}\frac{{d^3k}}{E^{\ast}(k)}
\frac{{\cal K}_{\mu\nu}+Q_{\mu\nu}M^{\ast 2}}{q^4-4(k\cdot q)^2}
\label{tt}
\eea
\vskip 1 true cm
\noindent where ${\cal K}_{\mu\nu}=(k_\mu-\frac{k.q}{q^2}q_\mu)
(k_\nu-\frac{k.q}{q^2}q_\nu)$, $Q_{\mu\nu}=(-g_{\mu\nu} +
\frac{q_{\mu}q_{\nu}}{q^2})$ and $E_k^*=\sqrt{{\bf k}^2+M^{\ast 2}}$. To include the overall degeneracy factor, the 
above expressions are multiplied by a factor of two coming from the nucleon
and proton loop.
It is clear that the form for the polarization tensor conforms to the
requirement of current conservation,
i.e.
\be
q^\mu\Pi^D_{\mu\nu}=0=\Pi^D_{\mu\nu}q^\nu .
\ee
In the present case we observe that the free part and the
dense part of the polarization tensor individually satisfies the above
condition.

We should also observe that Eq. (\ref{vttv}) is proportional to $Q_{\mu\nu}$
and therefore contribute equally to the longitudinal and transverse mode.
In fact, it is $K_{\mu\nu}$ which in matter, induces the splitting of these
two modes as we shall discuss later. Evidently, the Dirac part is also 
proportional to $Q_{\mu\nu}$ and therefore the modes remain degenerate on
account of Lorentz symmetry. It might be mentioned that at $|q|=0$ they
are degenerate in matter because of the rotational symmetry.

Also, it is worth to point out that we could describe these effects in the 
linear density approximation for low baryonic densities. In this approximation,
the term ${\cal K}_{\mu\nu}$ cannot contribute and therefore the longitudinal
and transverse part become degenerate and Eqs. (\ref{vv}-\ref{tt}) give
\bea
\Pi^{vv}_{T(L}&=&-4 g_v^2 \frac{M^* q_0^2}{q^4-4M^{*2}q_0^2} \rho_{B} \\
\label{ldenvv}
\Pi^{vt+tv}_{T(L)}&=& 4 g_v^2 (\frac{\kappa M^\ast}{4M}) 
\frac{q^4}{M^*(q^4-4M^{*2}q_0^2)} \rho_{B} \\
\Pi^{tt}_{T(L)}&=& 4 g_v^2 (\frac{\kappa}{2M})^2
\frac{M^* q^4}{q^4-4M^{*2}q_0^2} \rho_{B} . \label{ldens} \\
\eea

These results could also directly be obtained
by multiplying the forward scattering amplitude with the density. In the
present case we implemented the same by expanding the integrand 
up to second order or in other words schematically we retained terms
like $\int_0^{k_F} d^3k f_{\mu\nu}(q_0,|q|)(1+O(k))$.

To provide further insight, one can make a long wavelength approximation and
recover the familiar results of the plasma oscillations. In this limit,
when $q_0<E_F$ and $|q|<k_F$, Eq. (\ref{ldenvv}) reduces to 
$\Pi^{vv}_{T(L)}=g_v^2/M^* \rho_B$. In this limit, with $\kappa=0$ the
dispersion relation of the density dependent part alone becomes
\bea
q_0^2=|q|^2+m_v^2 + \Omega^2 ,
\eea
where the plasma frequency $\Omega^2=g_v^2/M^* \rho_B $. This is the
non-relativistic form of the results presented by Chin \cite{chin77} for
the case of $\omega$ meson propagation in nuclear matter. Furthermore,
replacing $g_v$ by the electronic charge ``e'' and putting $m_v=0$ what
one obtains is nothing but the familiar plasma frequency encountered in
condensed matter physics \cite{fetter}.

\begin{figure} [htb!]
\begin{center}
\epsfig{file=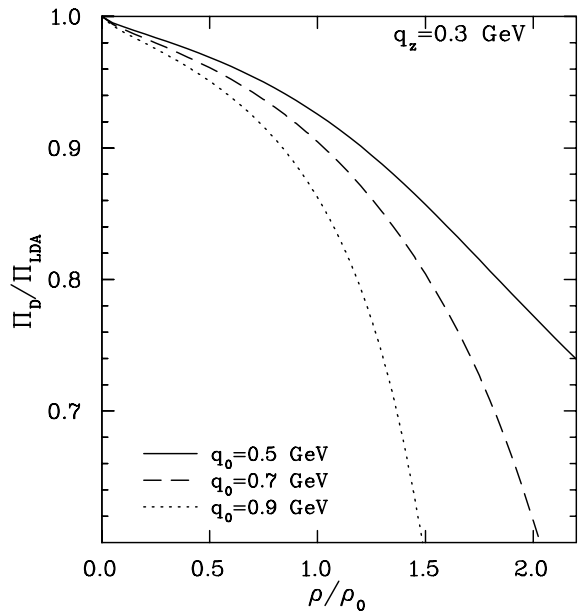,height=8.0cm,angle=0}
\end{center}
\caption{Comparison of the $\rho$-meson self-energy calculated at the one-loop 
level with the one calculated in the linear density approximation. 
\label{fig:LDA}}
\end{figure}

In Fig. (\ref{fig:LDA}) we compare resummed one loop results of the 
$\rho$-meson self-energy with the ones 
calculated in the linear density approximation. 
The ratio of the self-energies are shown as a function of density. 
It is now apparent that the linear approximation can be justified only up 
to normal nuclear matter density and we notice a rather strong dependence 
on energy, the approximation becoming worse for high values.

\subsection{Nucleon-Resonance Loop}

In the present work, among the resonances, we consider only $N^*(1520)$ and
$N^*(1720)$ which couples strongly with the $\rho$ meson as indicated in
\cite{peters98}. The corresponding relativistic interactions are given
by
\bea
\label{lagrangian}
{\cal L}_{int} = \left \{
\begin{array}{cclll}
\frac{f_{R N \rho}}{m_\rho}\,{\bar \psi}^{\mu} \, \gamma^\nu \, \psi
\, F_{\mu \nu} & \quad \mbox{for} & I(J^\pi) = \frac{1}{2}(\frac{3}{2}^-)\\ \\
\frac{f_{R N \rho}}{m_\rho}\,{\bar \psi}^{\mu} \, \gamma^5 \, \gamma^\nu \, \psi
\, F_{\mu \nu} & \quad \mbox{for} & I(J^\pi) = \frac{1}{2}(\frac{3}{2}^+)
\qquad .
\end{array} \right .
\eea
Here $\psi^\mu$ denotes the resonance spinor and $\psi$ the nucleon spinor,
$\sigma^{\mu\nu} = \frac{i}{2}\,\left[ \gamma^\mu , \gamma^\nu \right]$ and
$F^{\mu \nu} = \partial^\mu \, \rho^\nu -  \partial^\nu \, \rho^\mu$.
It might be mentioned that 
$N^*(1520)$ and $N^*(1720)$ have total widths of $\sim$ 120 MeV  and 150 MeV 
respectively
with corresponding branching ratios into the $\rho$-N channel of $\sim$
24 MeV  and 115 MeV .

\begin{figure} [htb!]
\begin{center}
\epsfig{file=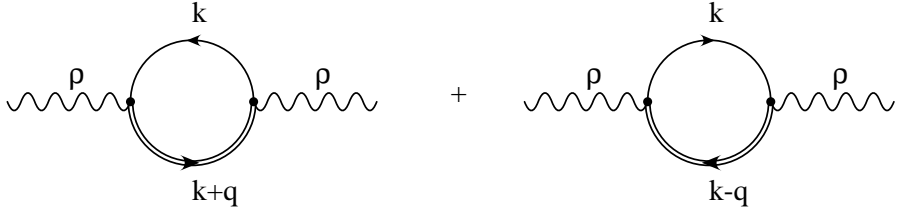,height=3.0cm,angle=0}
\end{center}
\caption{$\rho$ self-energy.} \label{fig:rho-self}
\end{figure}

From the Fig.\ref{fig:rho-self} it is evident that the polarization tensor 
has two
parts designated as direct and exchange term. We present only the direct
term and the exchange term can be calculated in a similar fashion.

\bea
-i\Pi_{\mu\nu}^{\pm(dir)} = S_I 
\int\frac{d^4k}{(2\pi)^4} 
\frac{T_{\mu\nu}^\pm(k,k-q)}{(k-q)^2-m_R^2}\big [
{\frac{1}{k^2-m_n^2} 
+ \frac{i\pi}{E_k}\delta(k^2-m_n^2)\theta(k_F-|k|)}\big ] ,
\label{pidir}
\eea
where $S_I$ is the isospin factor. 
The vertex factors for $R^{3/2}N\rho$ could be written as
\bea
\Gamma_{\mu\alpha}^\pm=\frac{f_\rho}{m_\rho}(\gamma_5)^{\frac{1\pm 1}{2}}
(\gamma_\mu q_\alpha - \qs g_{\mu\alpha})
\eea
  
It is evident that like the N-N loop, $\Pi_{\mu\nu}^{\pm(dir)}$
also contains a ``free'' and a density dependent part. The detailed 
expression for the free part is given in the appendix and has
a form $\Pi_{\mu\nu}=Q_{\mu,\nu} \Pi(q^2)$. Note also that
  
\bea
T_{\mu\nu}^\pm(k+q,k)&=&Tr[i(\ks+m_n)i\Gamma_{\mu\alpha}^\pm 
i{\cal R}_{3/2}^{\alpha\beta}(k-q)i\Gamma_{\beta,\nu}^\pm]\nn\\
&=&Tr[(\ks\mp m_n)(\gamma_\mu q_\alpha
                   -\qs g_{\mu\alpha})(\ks -\qs +m_R)\nn\\
&& \times P_{3/2}^{\alpha\beta}(k-q)
                     (q_\beta \gamma_\nu - \qs g_\beta\nu)] .
\eea
 
In the above equation $ {\cal R}^{\mu\nu}_{3/2}(p)$
is the Rarita-Schwinger propagator, given by
   \bea
{\cal R}^{\mu\nu}_{3/2}(p)
&=&(\ps + m_R) P^{\mu\nu}_{3/2}(p)\\
&=&(\ps + m_R) [-g^{\mu\nu}+\frac{1}{3}\gamma^\mu\gamma^\nu
 + \frac{2}{3}\frac{p^\mu p^\nu}{m_R^2}
-\frac{1}{3}\frac{p^\mu\gamma^\nu-\gamma^\mu p^\nu}{m_R}]
\eea
There is an overall sign ambiguity with
spin 3/2 particles which arises from the special choice of the point
transformation properties of the spin 3/2 Lagrangian 
\cite{banerjee95,ellis98}. This has also been discussed in 
Ref.\cite{post01}. However, we do not adopt here the prescription
suggested in Ref.\cite{post01} to deal with the spin 3/2 propagator.

The relevant trace can be written in the following suggestive form

\bea
T_{\mu\nu}^\pm(k,k-q)&=&(\frac{f_\rho}{m_\rho})^2\alpha^\pm 
Q_{\mu\nu}+ \beta^\pm K_{\mu\nu}
\eea
where,
\bea
\alpha^\pm &=&\frac{8}{3}
[-k^2q^2+m_n m_R q^2 + \frac{k^2q^4}{m_R^2}-2\frac{k^2q^2 (k\cdot q)}{m_R^2}
-\frac{q^4 (k\cdot q)}{m_R^2}+(k\cdot q)^2+\frac{k^2 (k\cdot q)^2}{m_R^2}
+2\frac{q^2 {k\cdot q}^2}{m_R^2}-\frac{(k\cdot q)^3}{m_R^2}]
 \nn \\ 
\beta^\pm &=& \frac{8}{3}( k\cdot q - k^2 -m_R^2)\frac{q^2}{m_R^2} .
\eea

It is evident that this structure is similar to what we had for the nucleon
loop and therefore satisfies the condition of current conservation
($q^\mu\Pi_{\mu\nu}=0=\Pi_{\mu\nu}q^\nu$ in the momentum space). 
$T^\pm_{\mu\nu}$ involves the same gauge invariant forms 
$K_{\mu\nu}$ and $Q_{\mu\nu}$.

In order to evaluate $\Pi^D_{\mu\nu}$ conveniently, we choose $\vec q$ to be
along the $z$ axis i.e. $ q=(q_0,0,0,\mid\vec q\mid) $, and
$k \cdot q=
E^{\ast}(k)q_0-|\vec k||q|\chi$,
where $\chi$ is the cosine of the
angle between $\vec k$ and $\vec q$. After $\phi$ integration the
non-vanishing components $\Pi^D_{\mu\nu}$ are 

\be
\pmatrix { \Pi_{00} & 0 & 0 & \Pi_{03} \cr
0 & \Pi_{11} & 0 & 0 \cr
0 &  0 & \Pi_{22} & 0 \cr
\Pi_{30} &  0 & 0 & \Pi_{33} } .
\ee
\vskip 0.5 true cm
Moreover for isotropic nuclear matter we have
$\Pi_{22}^D=\Pi^D_{33}$
and $\Pi^D_{01}=\Pi^D_{10}$, and
hence, taking all this into account, we have only two non-vanishing
independent components of $\Pi^D_{\mu\nu}$, linear combinations of
which gives us the longitudinal and transverse components of
$\Pi^D_{\mu\nu}$, in other words:
$\Pi^D_L(q)=-\Pi_{00}^D+\Pi_{33}^D$
 and
$\Pi_T^D(q)=\Pi^D_{11}=\Pi^D_ {22}$.

\begin{figure} [htb!]
\begin{center}\epsfig{file=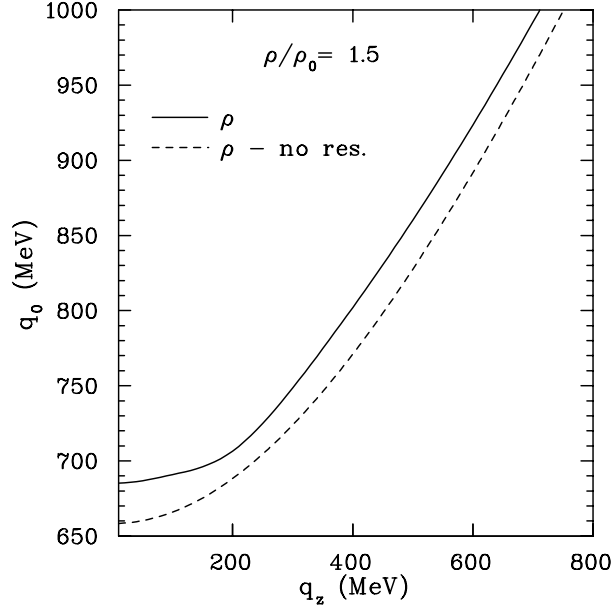,height=8.0cm,angle=0}
\end{center}
 \caption[Dispersion curve]
  {\small The dispersion curve for $\rho$ showing explicitly the 
effect of the baryonic resonances. See the text for details. 
\label{fig:disp}}
\end{figure}

We can now estimate the dispersion curves of the $\rho$ meson in nuclear 
matter. 
As mentioned, they appear as poles in the propagator and therefore zeros
of the dielectric functions shown in Eqs. (\ref{epsonT}) and (\ref{epsonL}). 
In Fig. \ref{fig:disp} we show the dispersion curves for $\rho$ meson for 
$\rho$=1.5$\rho_0$ baryonic density. It is clear that the $\rho$ meson
physical mass (when $q_z$=0) drops in nuclear matter from its free space 
value. The dashed curve shows the
results with n-n loop only and the solid one corresponds to the case where
the direct coupling of the $\rho$ with $N^*(1520)$ and $N^*(1720)$ is also
considered. It should be noted that the resonance-particle excitations
contribute with opposite sign to that of the n-n loop. This partially
offsets the dropping of $\rho$ meson mass in nuclear matter. 

\begin{figure} [htb!]
\begin{center}
\epsfig{file=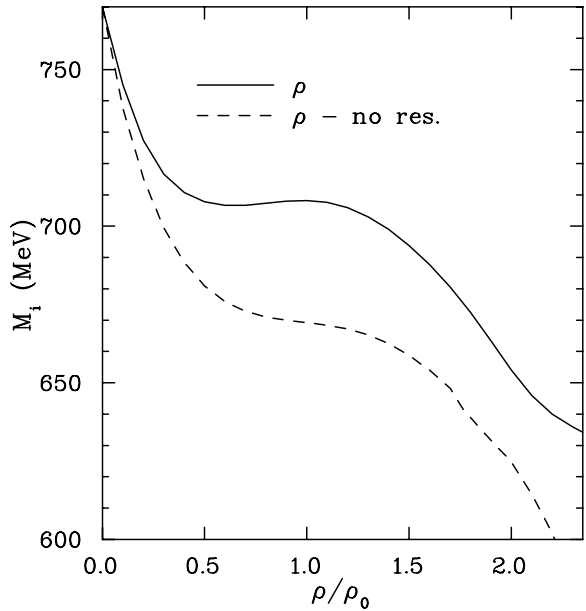,height=8.0cm,angle=0}
\end{center}
 \caption[Invariant Masses]
  {\small The invariant mass of the $\rho$ meson as a function
of density showing explicitly the effect of the baryonic resonances.
   \label{fig:imass}}
\end{figure}

Fig. \ref{fig:imass} shows the variation of the invariant mass of the
$\rho$ meson mass as a function of nuclear density. 
The Dirac vacuum and the density dependent part of the
self-energy contribute with opposite sign to the invariant mass. 
  
At lower
densities the free part is responsible for the dropping of the $\rho$ 
mass while at higher densities the masses again tend to increase
when the contribution from the density dependent part increases. This 
behaviour was also observed in the case of the $\omega$ and $\sigma$ mesons in
Ref. \cite{saito98}. 

\subsection {$\rho$-$a_0$ Mixing {\em via} N-N Loop}

Before we start the discussion on the $\rho$-$a_0$ mixing involving n-n 
polarization in nuclear matter, we should say a few words on the $a_0$ 
coupling to the nucleon. A more detailed study on the $a_0$ propagation in a 
dense medium could be found in \cite{teodorescu00}. The interaction 
can be described by the following Lagrangian:
\be
   {\cal L}_{int} = g_{a_0} {\bar \psi}\phi_{{a_0},a}\tau^a \psi
\ee
where $\psi$ and $\phi_{a_0}$ correspond
to the nucleon and $a_0$ fields, and $\tau_a$ 
is a  Pauli matrix. The values used for the coupling parameters are obtained 
from Ref. \cite{Bonn}. We do not involve the coupling of $a_0$ to the baryonic
resonances since currently this is not precisely known. 

\begin{figure} [htb!]
\begin{center}
\epsfig{file=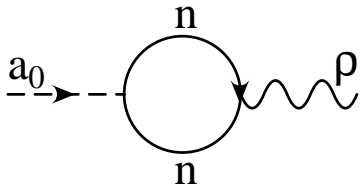,height=3.0cm,angle=0}
\end{center}
 \caption{ $a_0$-$\rho$ mixing via nucleon-nucleon loop.} \label{feynrhodel}
\end{figure}

The polarization vector through which the $a_0$ couples to $\rho$
via the n-n loop is obtained by evaluating the Feynmann diagram depicted 
in Fig. \ref{feynrhodel} and is given by

\bea
\Pi_ \mu (q_0,|{\vec q}|) &=& 2i g_{a_0} g_\rho \int \frac{d^4k}{(2\pi)^4}
    \mbox{Tr}[G(k) \Gamma_\mu G(k+q)] , \label{pim}
\eea

where 2 is an isospin factor. 
With the evaluation of the trace and after a little algebra Eq. (\ref{pim}) 
can be put into a suggestive form:
\bea
\Pi_\mu(q_0,|q|)=\frac{g_\rho g_{a_0}}{\pi^3} 2q^2 (2m_n^*-
\frac{\kappa q^2}{2 m_n})
\int_0^{k_F}\frac{d^3k}{E^*(k)}
\frac{k_\mu - \frac{q_\mu}{q^2}(k\cdot q)}{q^4 - 4 (k\cdot q)^2}\ .
\label{mixamp}
\eea
This immediately leads to two conclusions. First,
it respects the current conservation condition,
$q^\mu\Pi_\mu=0=\Pi_\nu q^\nu$. Secondly, there are only two components which
survive the integration over azimuthal angle. This
guarantees that it is only the longitudinal component of the $\rho$ meson
which couples to the scalar meson while the transverse
mode remains unaltered. Furthermore, current conservation implies that
out of the two non-zero components of $\Pi_\mu$, only one is independent.

In presence of mixing the combined meson propagator can be written
in a matrix form where the dressed propagator would no longer be a diagonal
matrix:
\bea
{\cal D} = {\cal D}^0 + {\cal D}^0\Pi{\cal D} .
\label{dressprop}
\eea
It is to be noted that the free propagator is diagonal having the following
form :
\be
{\cal D}^0 = \pmatrix{
    D^0_{\mu \nu} & 0\cr
    0 & \Delta_0
}
\label{free}
\ee
In Eq.(\ref{free}) the noninteracting propagator for $a_0$
and $\rho$ are given respectively by
\bea
\Delta_0(q) &=& \frac{1}{q^2 - m_{a_0}^2 + i\epsilon},
\label{frees} \\
D^0_{\mu \nu}(q) &=& \frac{-g_{\mu \nu}+ {\frac{q_\mu q_\nu}{q^2}}}{
  q^2 - m_\rho^2 + i\epsilon}, \label{freerho}
\eea
In fact, it is the polarization matrix which involves non-diagonal elements
as shown below characterizing the mixing
\be
\Pi = \pmatrix{
    \Pi^\rho_{\mu \nu}(q) & \Pi_\nu(q)\cr
    \Pi_\mu(q) & \Pi^{a_0}(q)}
\label{eq:off-diag}
\ee
After $\phi$ integration the
non-vanishing components $\Pi$ are as shown below   
\be
\pmatrix { \Pi_{00} & 0 & 0 & \Pi_{03} &\Pi_{0}\cr
0 & \Pi_{11} & 0 & 0 & 0\cr
0 &  0 & \Pi_{22} & 0 & 0\cr
\Pi_{30} &  0 & 0 & \Pi_{33} & \Pi_3 \cr
\Pi_{0} & 0 & 0 & \Pi_3 & \Pi^{a_0}}
\ee
\vskip 0.5 true cm

For $a_0$ meson the free part of the self-energy is
given by :
\bea
\Pi^{a_0}(q^2)=\frac{3 g_{a_0}^2}{2\pi^2}[3(m^{* 2}-m^2)
-4 (m^* - m)m
-(m^{* 2}-m^2)\int_0^1 dx\hspace{0.1cm} \ln \left[ \frac{m^{* 2} -
x(1-x)q^2}{m^2-x(1-x)q^2} \right] \nonumber\\
-\int_0^1 dx\hspace{0.1cm} (m^2-x(1-x)q^2)\ln \left[ \frac{m^{* 2} -
x(1-x) q^2}{m^2 - x(1-x)q^2} \right]\ .
\eea

To determine the collective modes, one defines the
dielectric function in presence of the mixed terms\cite{chin77}:
\bea
\epsilon(q_0,|{\vec q}|) &=& \det (1 - {\cal D}^0 \Pi)
   = \epsilon_T^2 \times \epsilon_{mix}
\eea
where $\epsilon_{T}$ corresponds to
two identical transverse (T) modes and $\epsilon_{mix}$ correspond to the
longitudinal mode with the mixing. The latter also characterizes
the mode relevant for the $a_0$ propagation
\bea
\epsilon_T &=& 1 - d_0 \Pi_T, \hspace{0.6cm}
d_0 = \frac{1}{q^2 - m_{{a_0}}^2 + i\epsilon}\label{trns} \nonumber\\
\epsilon_{mix} &=& (1 - d_0 \Pi_L)(1 - \Delta_0 \Pi_s) -
\frac{q^2}{|{\vec q}|^2} \Delta_0 d_0 (\Pi_0)^2 .
\label{long}
\eea

It is evident that only the longitudinal component gets modified because
of the mixing, and when $\Pi_0=0$ we recover the same expression as 
Eq. (\ref{epsonL}). 

\section{$\rho$ spectral density in nuclear matter}

Unlike in free space, the longitudinal and transverse 
$\rho$ spectral densities are non-degenerate. This is because of the fact
that, in the nuclear matter rest frame, Lorentz symmetry is broken and 
they are functions of $q_0$ and $|{\bf q}|$, independently. Furthermore, 
in matter the scalar and vector mesons can mix. This also
modifies the longitudinal $\rho$ spectral density through the off-diagonal
mixing terms in Eq. (\ref{eq:off-diag}).   

Now in the presence of mixing, the spectral densities can be defined 
in terms of the dielectric function as 
\bea
S_L(q_0,|{\bf q}|,\rho_B)=-\frac{1}{\pi} 
Im [\frac{d_0(1-\Delta_0\Pi_s)}{\epsilon_{SL}}] .
\eea
On the other hand the transverse spectral density is unaffected by the mixing
and has the following form 
\bea
S_T(q_0,|{\bf q}|,\rho_B)=-\frac{1}{\pi} Im [\frac{d_0}{1-d_0 \Pi_T}] ,
\eea

\begin{figure} [htb!]
\begin{center}
\epsfig{file=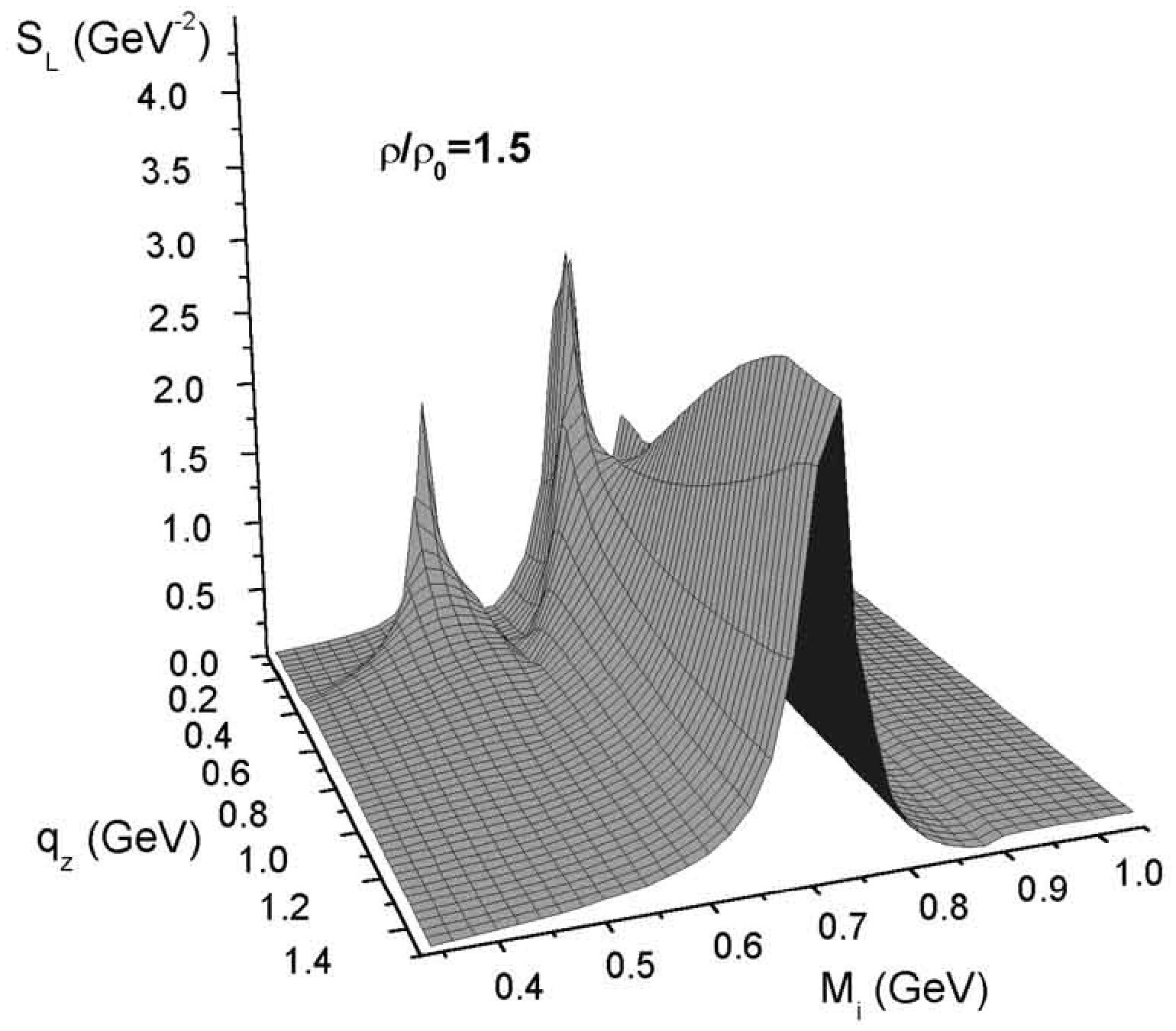,height=7.0cm,angle=0}

\epsfig{file=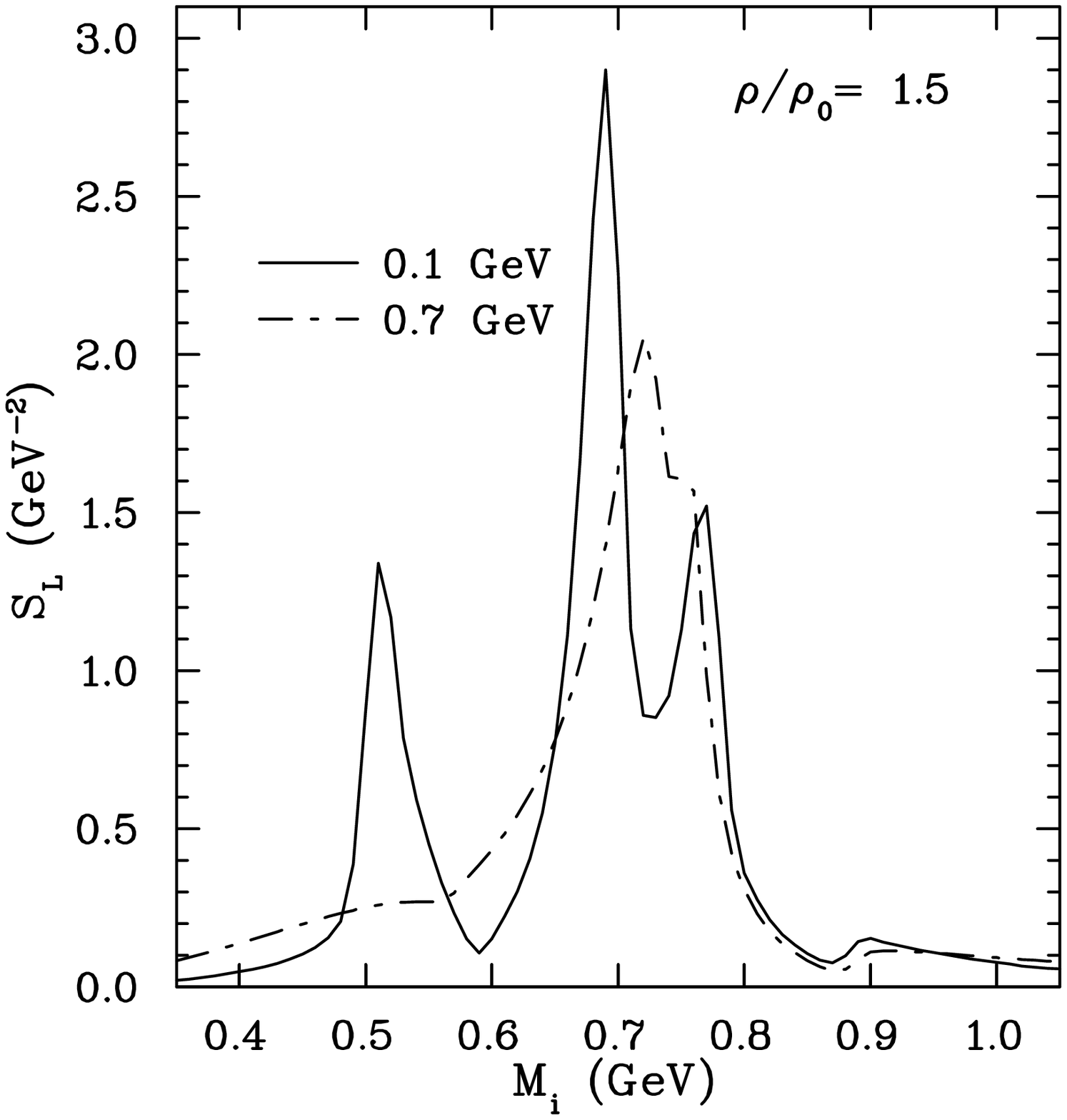,height=6.0cm,angle=0}
\end{center}
 \caption{The longitudinal spectral density for the $\rho$ meson with mixing
at density $\rho=1.5 \rho_0$.
}
\label{specL15}
\end{figure}
 
Fig.\ref{specL15} shows the longitudinal spectral density ($S_L$) at a density
$\rho=1.5 \rho_0$ as a function of three momenta ($|{\bf q}|$=$q_z$) 
and invariant mass (M$_i$). 
This includes the effect of N-N loop and the direct coupling of the
$\rho$ meson with $N^*(1520)$ and $N^*(1720)$. The three-peak structure
of the spectral density is similar to what has already been observed in
non-relativistic calculations in Ref. \cite{urban98}. Such a 
characteristic behaviour of the spectral density in presence of resonance 
has been also observed in pion-nucleon dynamics. The collective modes 
induced by 
the density fluctuations can be identified as the $\rho$ meson mode, N-hole 
and N-resonance mode similar to what we observe in case of pion propagation in
nuclear matter. 

\begin{figure} [htb!]
\begin{center}
\epsfig{file=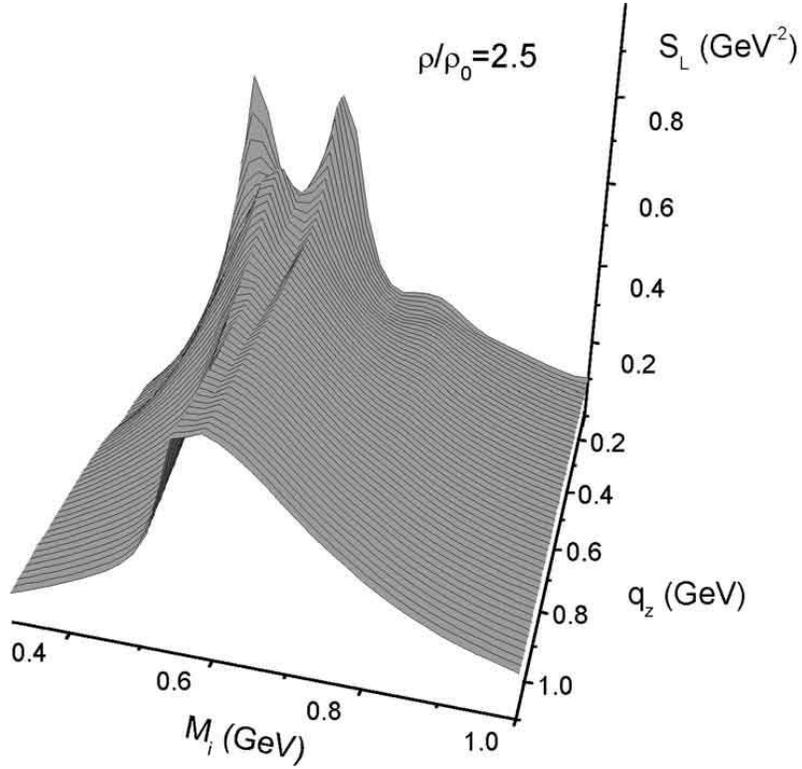,height=11.0cm,angle=0}
\end{center}
 \caption{
The longitudinal spectral density for the $\rho$ meson with mixing
at density $\rho=2.5 \rho_0$.
}
\label{specL25}
\end{figure}
Fig. \ref{specL25} is as Fig. \ref{specL15} but at higher density. 
We note that at higher density the spectral density gets even broader. 
Furthermore,
with increasing momenta all the peaks merge into one broad peak indicating
the fact that at high momenta the collective behaviour dies down and the
meson shows a behaviour of a free propagation in matter. 

\begin{figure} [htb!]
\begin{center}
\epsfig{file=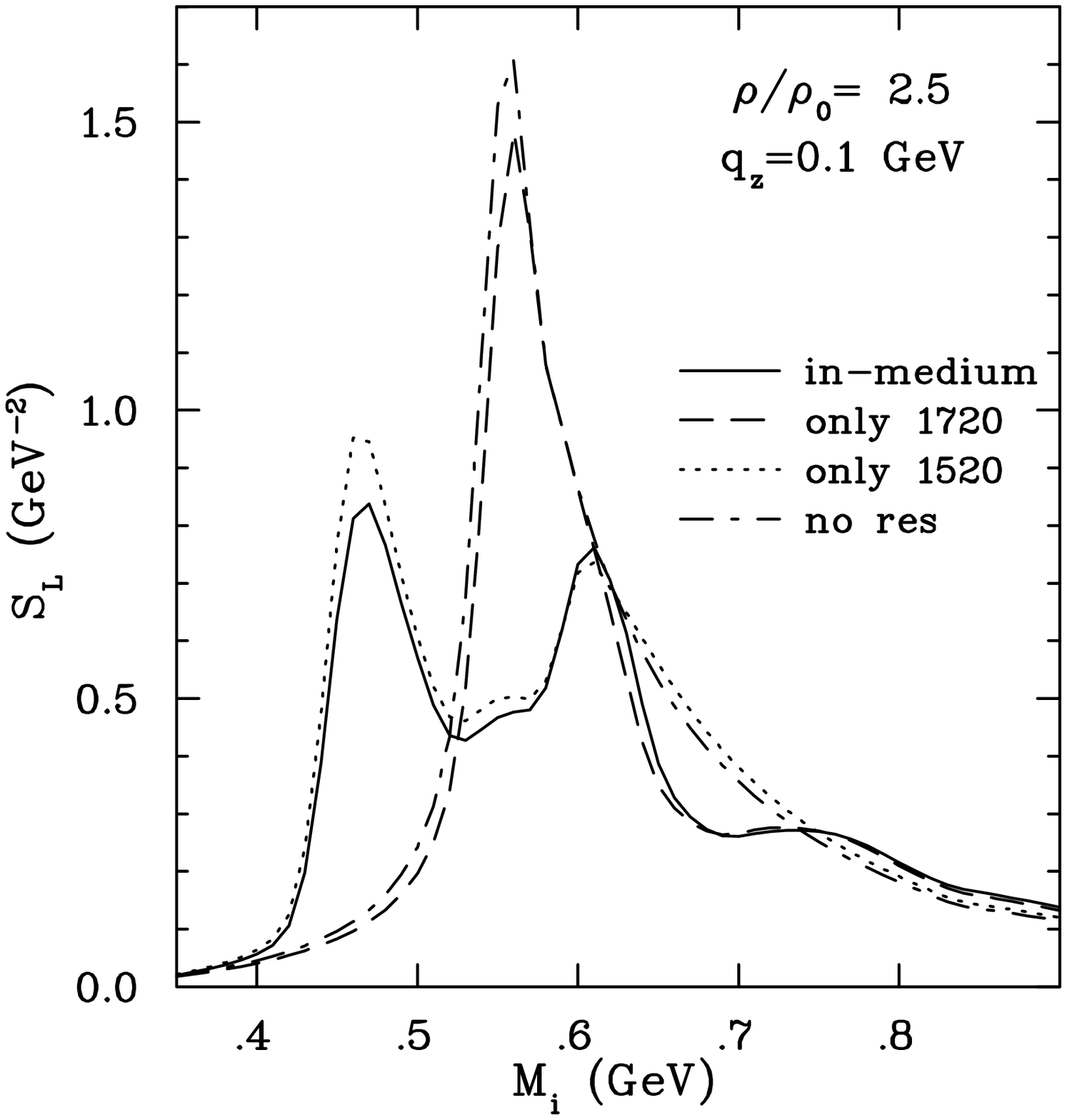,height=6.0cm,angle=0}\epsfig{file=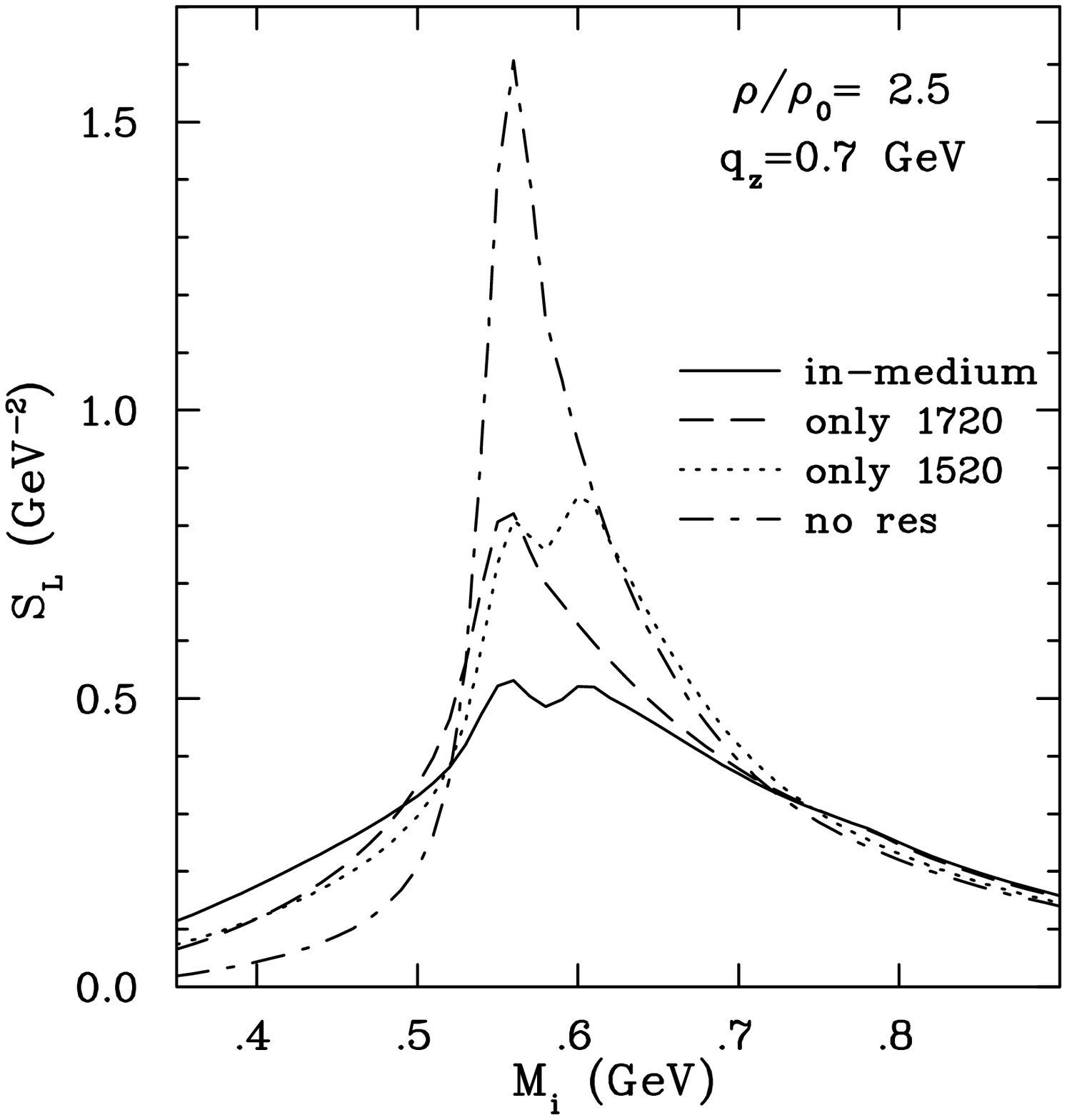,height=6.0cm,angle=0}
\end{center}
 \caption{
The longitudinal spectral density for the $\rho$ meson with mixing
at density $\rho=2.5 \rho_0$ with and without considering resonances.
}
\label{specind}
\end{figure}

Next we show how individual resonant states modify the $\rho$ spectral
density in matter. In Fig. \ref{specind}, the dashed-dotted line represent
the $\rho$ spectral density for $\rho$ coupled only to $\pi$-$\pi$ and
N-N loop. This shows the $\rho$ meson peak shifts towards lower invariant 
mass.

\begin{figure} [htb!]
\begin{center}
\epsfig{file=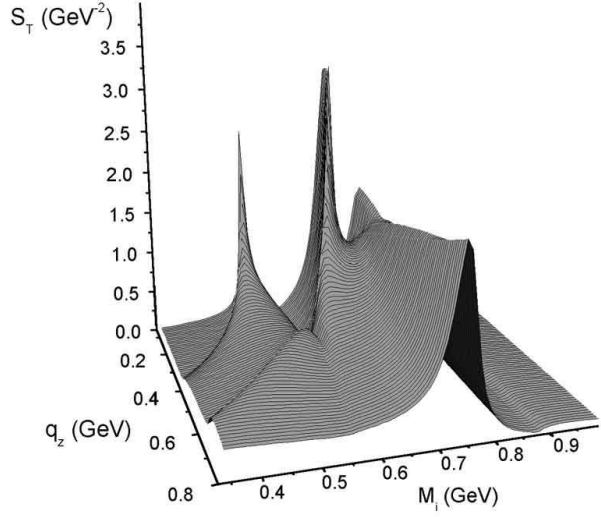,height=8.0cm,angle=0}
\end{center}
 \caption{
The transverse spectral density for the $\rho$ meson with mixing
at density $\rho=1.5 \rho_0$.
}
\label{specT15}
\end{figure}
This is because of the $\sigma$ meson mean field in nuclear matter which
lowers the nucleon mass considerably in nuclear matter. The dashed curve
has a double hump structure with the introduction of the $N^*(1720)$. For
non-relativistic calculation, such a feature has already been observed in
the static limit in Ref. \cite{friman}. Addition of $N^*(1520)$ also introduces
a double peak structure but it moves the strength to the low invariant mass
region. Hence we conclude that the resonant states are responsible for the
broadening of the $\rho$ spectral density in nuclear matter. Furthermore,
in presence of the resonant states the $\rho$ meson gets broadened so much 
so that the single particle interpretation of $\rho$ meson as quasi particle
fails to carry any sense.

The transverse spectral density ($S_T$) also shows interesting features at 
low momenta. Again we observe that at higher momenta the $\rho$-spectral
density gets flattened. The momentum dependence of $S_T$
is also observed to be different than that of $S_L$.
\begin{figure} [htb!]
\begin{center}
\epsfig{file=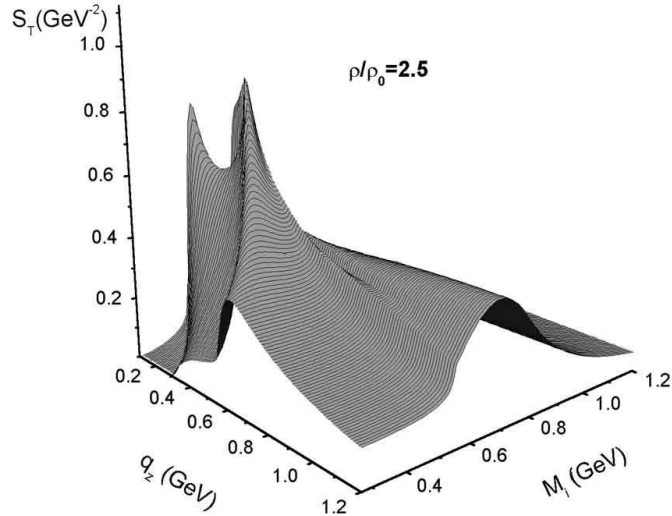,height=8.0cm,angle=0}
\end{center}
 \caption{
The transverse spectral density for the $\rho$ meson with mixing
at density $\rho=2.5 \rho_0$.
}
\label{specT25}
\end{figure}

At higher densities we see even more pronounced effect as evident from
Fig. \ref{specT25}. 

The spectral density for the $a_0$ meson is defined as:
\bea
S_S(q_0,|{\bf q}|,\rho_B)=-\frac{1}{\pi} 
Im [\frac{\Delta_0(1-d_0\Pi_s)}{\epsilon_{SL}}] .
\eea
\begin{figure} [htb!]
\begin{center}
\epsfig{file=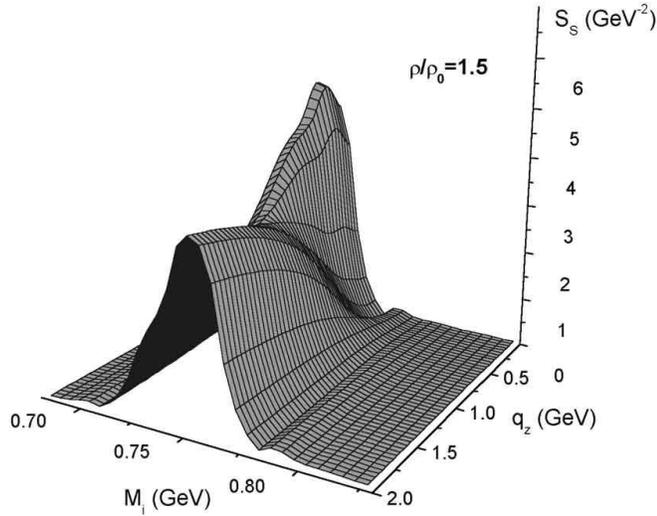,height=8.0cm,angle=0}
\end{center}
 \caption{
Spectral density for the $a_0$ meson with mixing
at density $\rho=1.5 \rho_0$.
}
\label{speca015}
\end{figure}

\begin{figure} [htb!]
\begin{center}
\epsfig{file=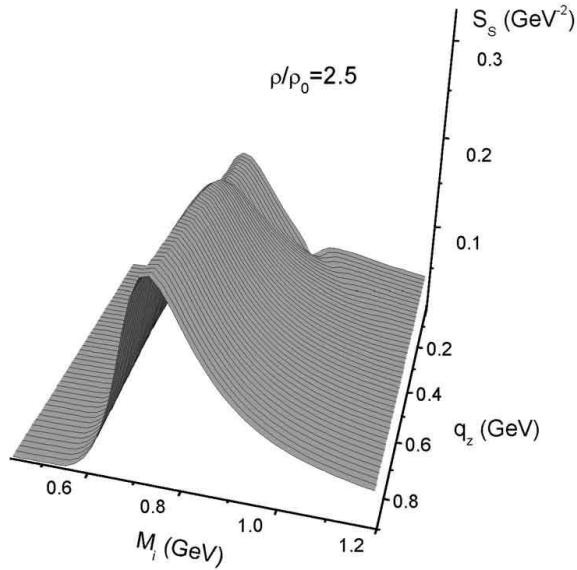,height=8.0cm,angle=0}
\end{center}
 \caption{
Spectral density for the $a_0$ meson with mixing
at density $\rho=2.5 \rho_0$.
}
\label{speca025}
\end{figure}

In Figs. \ref{speca015} and \ref{speca025} the spectral densities of the 
$a_0$ meson
are presented at densities $\rho=1.5 \rho_0$ and $\rho=2.5\rho_0$. It
is evident that the $a_0$ spectral density also gets broadened in
medium. A marked shift towards the lower invariant mass indicates that
the $a_0$ mass also drops in nuclear matter.

\begin{figure} [htb!]
\begin{center}
\epsfig{file=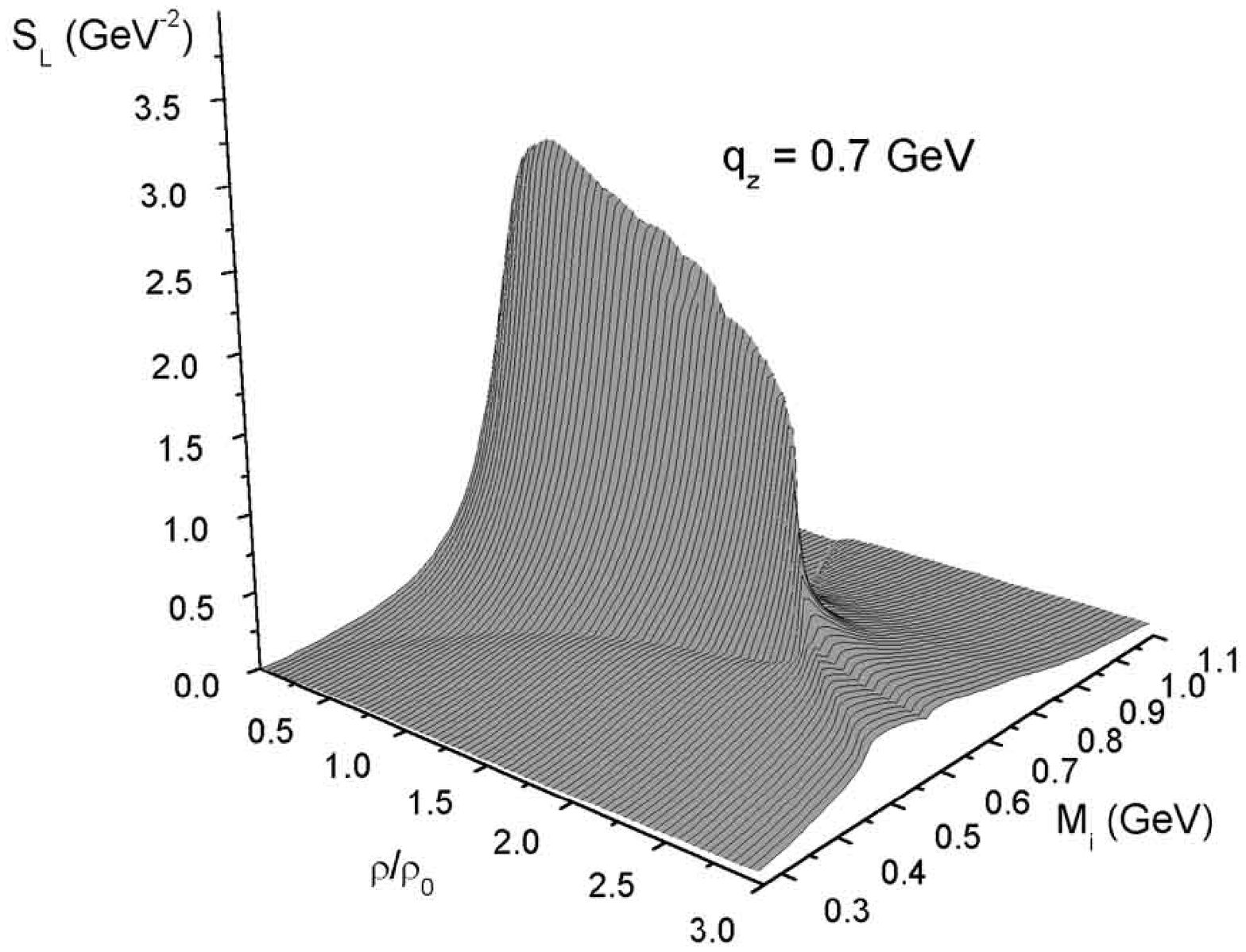,height=8.0cm,angle=0}

\epsfig{file=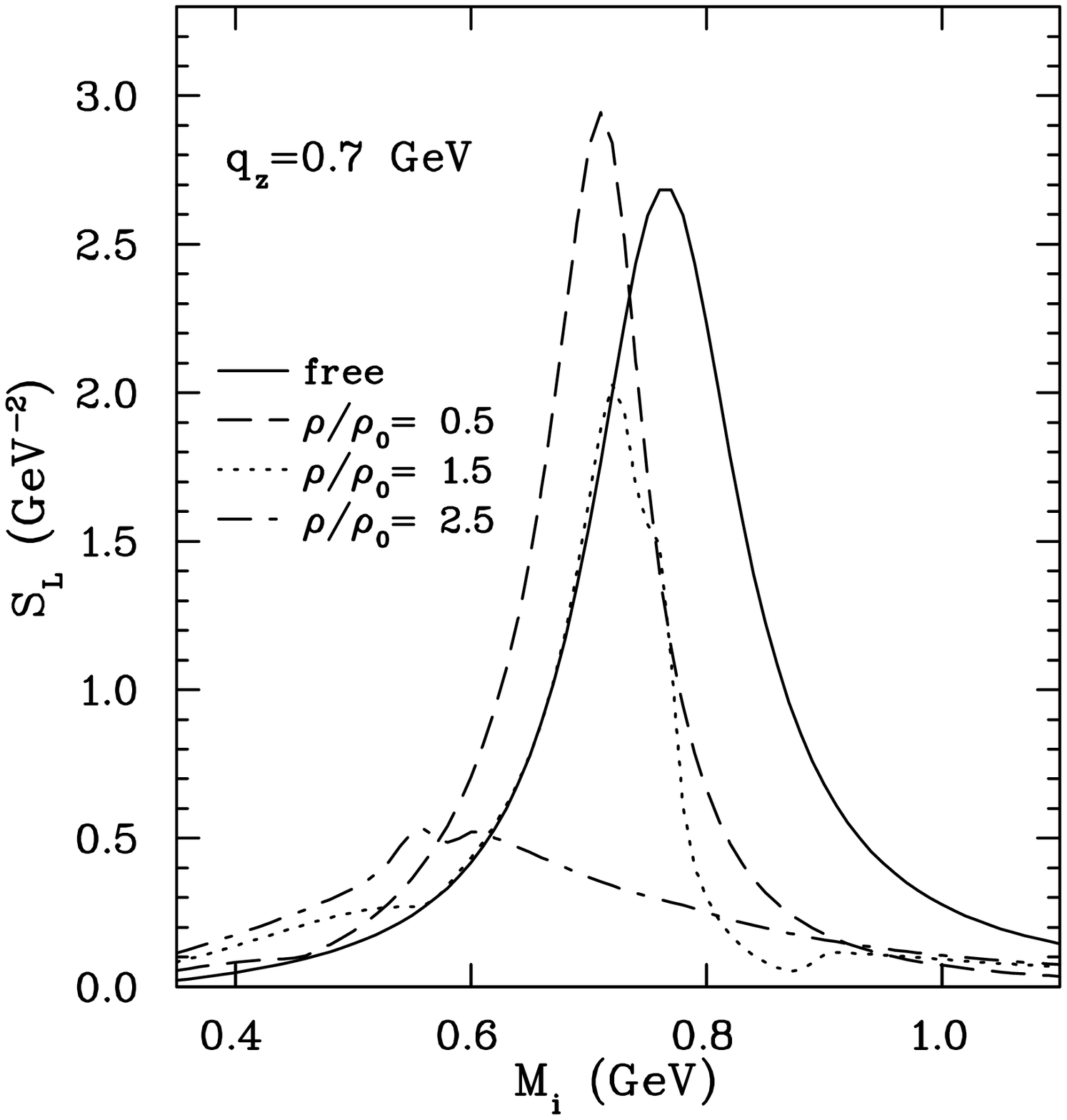,height=7.7cm,angle=0}
\end{center}
 \caption{The longitudinal spectral density for the $\rho$ meson with mixing
as a function of density.
}
\label{specLmix}
\end{figure}

To illustrate the spectral density modification in matter, we present
$S_L$ as a function of density. The solid line in Fig.\ref{specLmix}
shows the free $\rho$ spectral density. We find that with increasing
density it acquires more strength in the low invariant mass region and
also becomes flattened. Dashed, dotted and dashed-dotted lines represent
results for 0.5, 1.5 and 2.5 normal nuclear matter densities. It might
be mentioned that a small peak appears at the higher invariant mass region 
which indicates the effect of mixing.

\begin{figure} [htb!]
\begin{center}
\epsfig{file=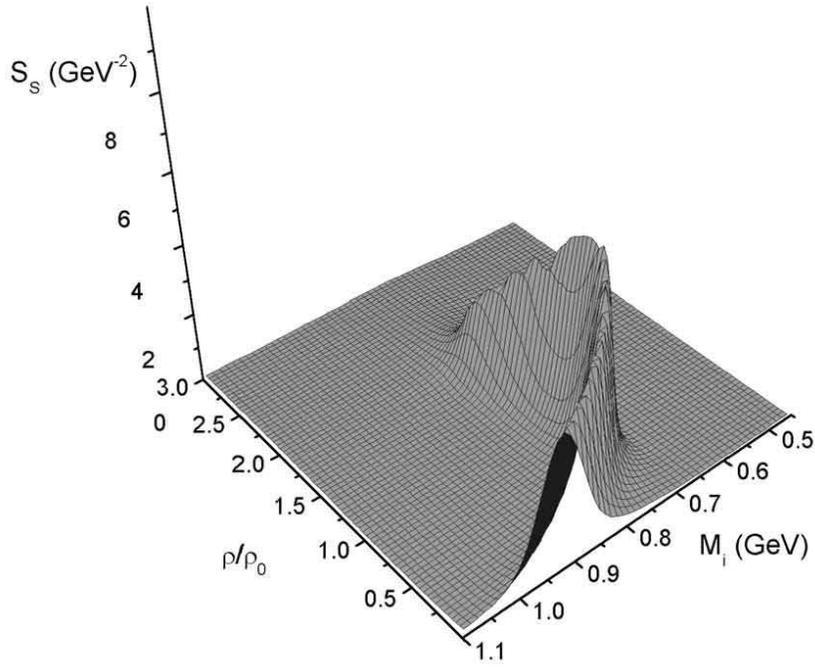,height=10cm,angle=0}
\vspace*{3mm}

\epsfig{file=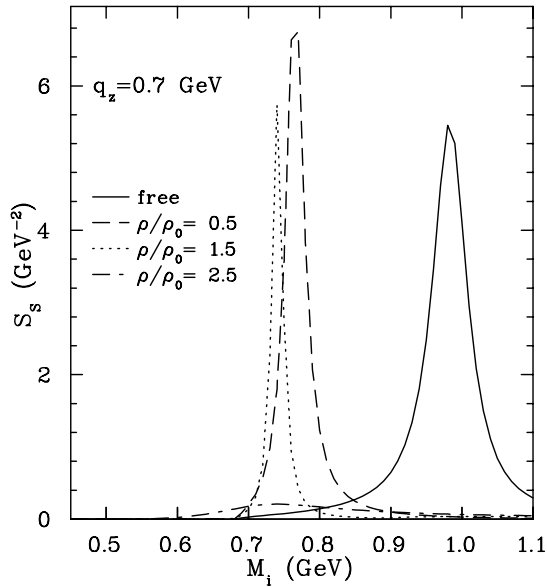,height=7.7cm,angle=0}
\end{center}
 \caption{The spectral density for the $a_0$ meson with mixing
as function of density (upper part) and a cross section for different densities (lower part).
}
\label{specSmix}
\end{figure}

In Fig.\ref{specSmix}, the two dimensional projection of the $a_0$
spectral density at $|{\bf q}|=0.7$ shows that with increasing density 
$a_0$ gets narrower and moves towards low invariant mass region. This
is understandable from the reduction of phase space. It might be 
recalled here that resonant states do not couple to $a_0$ and therefore
we do not observe any broadening in the $a_0$ channel.

\section{Summary and Conclusions}

In the present work we present quantitative results of the in-medium 
meson properties. The meson spectral densities are evaluated for the first 
time in a fully relativistic mean field model which goes beyond linear density 
approximation and we discuss the limitation of the aplicability of LDA. 
We find that meson spectral densities in matter are quite different than in 
free space. 
The difference stems partly from the Lorentz symmetry breaking. 
While in free space the transverse and longitudinal
component of the $\rho$ spectral density are degenerate, in matter, they
show different qualitative behaviour. Furthermore, we include other purely 
in-medium effects, forbidden in vacuum on the account of Lorentz symmetry, like
meson mixing. In matter the $\rho$
can also be modified because of its mixing with a scalar (isovector) $a_0$
meson. This effect, as we have seen, modifies only the longitudinal component
of the spectral density. 

In our model we observe that the $\rho$ meson spectral density gets 
flattened in nuclear matter with the incorporation of the resonant states
like $N^*(1520)$ and $N^*(1720)$. In fact in nuclear matter the original
free space peak of the $\rho$ meson become so broadened that it is no
longer possible to interpret that as a quasi-particle excitation. This
was also observed in Ref. \cite{peters98,post01}. It should also be noted
that in presence of scalar mean field the $\rho$ meson mass goes down as
a function of density.

This broadened of $\rho$ spectral density shows an accumulation of strength
towards lower invariant mass region. This would definitely imply that in
more production of dilepton pairs with low invariant mass in access to what 
we might expect from the free $\rho$ spectral density. We plan to extend this 
work to the finite temperature region in the near future. Furthermore, the
broadened $\rho$ spectral density would also affect the width of the
resonance states like $N^*(1520)$ or $N^*(1720)$ which probably would shed
some light on the issue of mixing resonant states in the photoabsorption
cross-sections. For the complete determination of the $\rho$ spectral 
density and the broadening of the resonant state, a self-consistent approach
should be adopted \cite{post01}. Studies along such directions are in progress.
\acknowledgements

This work was supported in part by the Natural Sciences and Engineering
Research Council of Canada and in part by the Fonds FCAR of the
Qu\'ebec Government.

\appendix
\section{Free part of the N-N Loop}
 Polarization tensors arising out of the $n-\bar n$ excitation of the Dirac sea.
\bea
\Pi_{\mu\nu}^{vv}=-\frac{1}{2}(\frac{g_v}{\pi})^2[\frac{1}{3}
(\Delta + ln\mu^2)q^2Q_{\mu\nu} -
2q^2Q_{\mu\nu}\int_0^1 dx x(1-x)lnD]
\eea

\bea
\Pi_{\mu\nu}^{vt+tv} = \frac{1}{2}\frac{g_v^2}{\pi^2}\frac{M^\ast\kappa_v}{2 M}
Q_{\mu\nu}[\Delta + ln\mu^2 -\int_0^1dx lnD]
\eea

\bea
\Pi_{\mu\nu}^{tt}=-\frac{g_\rho^2}{(4\pi)^2} (\frac{\kappa}{M})^2
Q_{\mu\nu}[\frac{1}{6}q^2 ( \Delta + m^2 ln(\mu^2) - m^2 + \frac{q^2}{q^2}
+ q^2\int_0^1 dx x^2 lnD -q^2 \int_0^1 dx x lnD - m^2 \int_0^1 dx lnD )
\eea

$$\Delta = \frac{1}{\epsilon} -\gamma + ln4\pi$$
$$ D= M^{\ast 2} -q^2 x(1-x)$$

All the terms containing $\Delta$ are infinite and need to be subtracted out.

\section{Free part of the N-R Loop}

\bea
\Pi_{\mu\nu}^{RN (dir)}=Q_{\mu\nu}\Pi(q^2)\nn\\
\eea
where,
\bea
\Pi(q^2,\mu)=
(\Pi_1(q^2)+
\Pi_2(q^2)+
\Pi_3(q^2)+
\Pi_4(q^2)+
\Pi_5(q^2)+
\Pi_6(q^2)+
\Pi_7(q^2)+
\Pi_8(q^2)
)
\eea

{
\bea
\Pi_1(q^2)&=&[ (D+2x^2q^2)(\Delta+\ln m^2 ) +
\int_0^1 dx (D - D\ln D - 2q^2x^2\ln D)] \frac{q^2Q_{\mu\nu}}{32\pi^2}\\ \nn 
\Pi_2(q^2)&=&[ 2(3D^2+18Dq^2x^2+4q^4x^4)(\Delta+\ln \mu^2) + \\ \nn
&+&\int_0^1 dx (9D^2+36Dq^2x^2+6D^2\ln D -36Dq^2x^2\ln D-8q^4x^4\ln D)]
\frac{q^2Q_{\mu\nu}}{128m_R^2\pi^2} \\\nn
\Pi_3(q^2)&=&-[2q^2x(3D+2q^2x^2)(\Delta+\ln \mu^2)+\\\nn
&+&\int_0^1 dx (3q^2xD-3q^2xD\ln D-2q^4x^3\ln D)]
\frac{q^2Q_{\mu\nu}}{32m_R^2\pi^2}\\\nn
\Pi_4(q^2)&=-&[(2D+q^2x^2)(\Delta+\ln \mu^2)+\\\nn
&+&\int_0^1 dx (D-2D\ln D-q^2x^2\ln D)]
\frac{q^2Q_{\mu\nu}}{16\pi^2}\\\nn
\Pi_5(q^2)&=&-m_Nm_\Delta[(\Delta+\ln \mu^2)-\int_0^1 dx \ln D]
\frac{q^2Q_{\mu\nu}}{16\pi^2}\\\nn
\Pi_6(q^2)&=&[D(\Delta+\ln \mu^2)+\int_0^1 dx(D-D\ln D)]
\frac{q^2Q_{\mu\nu}}{32\pi^2}\\\nn
\Pi_7(q^2)&=&[2D(3D+2q^2x^2)(\Delta+\ln \mu^2)+\int_0^1 \frac{dx}{m_R^2}
D(9D+4q^2x^2-6D\ln D -4q^2x^2\ln D)]\frac{q^2Q_{\mu\nu}}{128\pi^2}\\\nn
\Pi_8(q^2)&=&-[\int_0^1 dx Dq^2x(\Delta+\ln \mu^2)+\int_0^1 \frac{dx}{m_R^2} 
(Dq^2x-q^2xD\ln D)]\frac{q^2Q_{\mu\nu}}{32\pi^2}\\\nn
\eea
}
   
\section{Identities involving Rarita-Schwinger spinors}

\bea
\Delta^{\mu\nu}(p)&=&\sum \Psi^\mu(p){\bar\Psi}^\nu(p)\nonumber\\
&=&(\ps + \mr)
[-g^{\mu\nu}+\frac{1}{3}\gamma^\mu\gamma^\nu
 + \frac{2}{3}\frac{p^\mu p^\nu}{\mr^2}
-\frac{1}{3}\frac{p^\mu\gamma^\nu-\gamma^\nu p^\mu}{\mr}]\nonumber\\
&=&(\ps + \mr)P^{\mu\nu}_{3/2}(p)
\eea
\bea
\gamma_\mu P^{\mu\nu}_{3/2}(p)=
\frac{1}{3\mr}(\mr^2 - p^2)(\gamma_\nu - \frac{2 p_\nu}{\mr})
\eea
Clearly the above and the following equations vanish when the spin 3/2 
state  is
on-shell, so does Eq. (C3) also.
\bea
P^{\mu\nu}_{3/2}(p)\gamma_\nu=
\frac{1}{3\mr}(\mr^2 - p^2)(\gamma_\mu - \frac{2 p_\mu}{\mr})
\eea
Another useful relation in this respect is the identity
\bea
(k - p)_\mu P^{\mu\nu}_{3/2}(p) (k -p)_\nu
= \frac{2}{3}\frac{(p^2-\mr^2)}{\mr^2}[p^2 - 2 k\cdot p] +
\frac{2}{3\mr^2}[ (k\cdot p)^2 - \mr^2 k^2]
\eea
It is to be noted that the above equation also takes much simpler form
when we have on-shell spin 3/2 projection operator.
\bea
p_\mu \Psi^\mu(p) = 0 = \Psi^\nu(p) p_\nu\\
k_\mu P^{\mu\nu}_{3/2}(p)
k_\nu = \frac{2}{3\mr^2}[ (k\cdot p)^2 - \mr^2 k^2]
\eea
It is evident that $p_\mu P^{\mu\nu}_{3/2}(p) p_\nu = 0 $  when
the spin 3/2 particle is on-shell {\it i.e.} $p^2=\mr^2$.
In the rest frame of the resonant state R, 
we, therefore, have $k_\mu P^{\mu\nu}_{3/2} k_\nu = \frac{2}{3}\kb^2$.

\end{document}